\documentclass[a4paper,11pt]{article}
\pdfoutput=1 
\usepackage{jheppub} 
\usepackage{epsfig}
\usepackage{caption}
\usepackage{subcaption}
\usepackage{color,soul}
\usepackage{float}
\usepackage{amsfonts}
\usepackage{url}
\usepackage{slashed}
\usepackage{accents}
\usepackage{lipsum}
\usepackage{mathtools}
\usepackage{physics}

\usepackage[normalem]{ulem}

\hypersetup{
  bookmarks=true,         
  unicode=false,          
  pdftoolbar=true,        
 pdfmenubar=true,        
 pdffitwindow=true,     
 pdfstartview={FitH},    
 pdfsubject={Dark Matter},   
 pdfnewwindow=true,      
 pdfcreator={RevTeX},
 colorlinks=true,       
 linkcolor=red,          
 citecolor=blue,        
 filecolor=black,      
 urlcolor=blue,           
  }

\title{Cosmological implications of inflaton-mediated dark and visible matter scatterings after reheating}
\author[a]{Deep Ghosh,}
\author[b]{Sourav Gope}
\author[b]{and Satyanarayan Mukhopadhyay}
\affiliation[a]{Department of Physical Sciences, Indian Institute of Science Education and Research (IISER) Kolkata, Campus Road, Mohanpur, West Bengal 741246.}
\affiliation[b]{School of Physical Sciences, Indian Association for the Cultivation of Science, 2A and 2B Raja S.C. Mullick Road, Kolkata 700 032}

\emailAdd{matrideb1@gmail.com}
\emailAdd{intsg5@iacs.res.in}
\emailAdd{tpsnm@iacs.res.in}

\abstract{The initial density of dark matter (DM) particles, otherwise secluded from the standard model (SM), may be generated at reheating, with an initial temperature ratio for internal thermalizations, $\xi_i=T_{\rm DM,i}/T_{\rm SM,i}$. This scenario necessarily implies inflaton-mediated scatterings between DM and SM after reheating, with a rate  fixed by the relic abundance of DM and the reheat temperature. These scatterings can be important for an inflaton mass and reheat temperature as high as $\mathcal{O}(10^7 {~\rm GeV})$ and $\mathcal{O}(10^9{~\rm GeV})$, respectively, since the thermally averaged collision terms become approximately independent of the inflaton mass  when the bath temperature is larger than the mass. The impact of these scatterings on DM cosmology is studied modeling the perturbative reheating physics by a gauge-invariant set of inflaton interactions upto dimension-5 with the SM gauge bosons, fermions and the Higgs fields. It is observed that an initially lower (higher) DM temperature will rapidly increase (decrease), even with very small couplings to the inflaton. There is a sharp lower bound on the DM mass below which the relic abundance cannot be satisfied due to faster  back-scatterings depleting DM quanta to SM particles. For low DM masses, the CMB constraints become stronger due to the collisions for $\xi_i<1$, probing values as small as $\mathcal{O}(10^{-4})$, and weaker for $\xi_i>1$. The BBN constraints become stronger due to the collisions for lower DM masses, probing $\xi_i$ as small as $\mathcal{O}(0.1)$, and weaker for higher DM mass. Thus inflaton-mediated collisions with predictable rates, relevant even for high-scale inflation models, can significantly impact the cosmology of light DM.}

\begin{document} 
\maketitle
\flushbottom  
\section{Introduction and Summary}
\label{sec:sec1}
Cosmic inflation is a candidate theory for the pre-radiation dominated epoch of the Universe. To begin with, it addresses the question of why the cosmic microwave background (CMB) has such a uniform temperature beyond the region which could have been causally connected with standard radiation and matter domination history. At the same time, the nearly scale invariant Gaussian density fluctuations imprinted in the CMB find a natural explanation in the quantum fluctuations of the inflaton field~\cite{Weinberg:2008zzc,Starobinsky:1979ty, Starobinsky:1980te, Kazanas:1980tx, Guth:1980zm,Linde:1981mu,Linde:1982uu,Albrecht:1982wi}.

Most models of inflation feature a scalar singlet field as the inflaton, which can, in general, couple to the dark matter (DM) and standard model (SM) fields. At least some of these couplings are necessary for reheating to a radiation dominated epoch in which the big-bang nucleosynthesis (BBN) can take place. A simple model for the reheating phase considers the damped oscillation of the inflaton field around the minimum of its potential after the end of the slow-roll epoch, where the potential can be approximated by a quadratic form near the minimum. Such an oscillatory phase is equivalent to the field theory of spin-0 particles with negligible velocity. This perturbative reheating phenomenon has been studied in detail, see, for example, Refs.~\cite{Weinberg:2008zzc,Linde:1981mu,Albrecht:1982mp,Dolgov:1982th,Abbott:1982hn,  Dolgov:1989us, Traschen:1990sw,Kofman:1994rk,Kofman:1997yn,Allahverdi:2010xz,Garcia:2023eol,Kanemura:2023pnv}. The possibility of non-thermal production of DM in such perturbative reheating is also well known~\cite{Moroi:1994rs, Kawasaki:1995cy, Moroi:1999zb, Chung:1998zb, Chung:1998ua, Jeong:2011sg, Ellis:2015jpg, Harigaya:2014waa, Drees:2021lbm, Garcia:2018wtq, Harigaya:2019tzu, Garcia:2020eof,  Garcia:2020wiy, Moroi:2020has,Adshead:2016xxj,Hardy:2017wkr,Berlin:2016gtr,Adshead:2019uwj,March-Russell:2020nun, Bhattacharya:2020zap, Haque:2022kez, Haque:2021mab, Ghosh:2022hen, Garcia:2022vwm, Ghoshal:2022jeo, Kaneta:2022gug, Ghoshal:2022ruy}.

In the conventional treatment of DM production in perturbative reheating, one usually computes the DM momentum distribution  at a given epoch by simply red-shifting the momenta at production, which for a two-body decay of a heavy inflaton is $p \sim m_{\rm inflaton}/2$. However, as shown by some of the present authors in Ref.~\cite{Ghosh:2022hen}, in such scenarios the DM necessarily undergoes both $s-$channel and $t-$channel scattering and annihilation processes involving the SM particles, mediated by the inflaton field. Therefore, if the inflaton is not very heavy compared to the reheat temperature, these scatterings can become significant in modifying the momentum distribution of DM. In fact, compared to the simplistic treatment of only red-shifting the momenta, the average DM velocity at the epoch of matter-radiation equality, and hence its free-streaming length, can be modified by more than an order of magnitude due to these scatterings, leading to significant weakening of the Lyman-alpha constraints on light DM~\cite{Kolb:1990vq, Boyarsky:2008xj, Irsic:2017ixq, Ghosh:2022hen}. 

In Ref.~\cite{Ghosh:2022hen}, we focused on the case of a singlet fermion DM, which remained non-thermal throughout its entire thermal history. On the other hand, if one considers a scalar singlet DM, the scalar singlet can naturally possess both elastic and number-changing self-interactions to internally thermalize~\cite{Davidson:2000er}. Thus we can have an internally thermalized DM with its own temperature, which in general is different from the SM temperature~\cite{Adshead:2016xxj,Hardy:2017wkr,Berlin:2016gtr,Adshead:2019uwj,March-Russell:2020nun}. In this paper, we investigate the role of the inflaton-mediated dark matter and standard model particle scattering processes in modifying the DM cosmology, by taking into account all the possible gauge-invariant couplings of the inflaton with the SM gauge, fermion and Higgs fields. For other related interesting ideas on the reheating and temperature of a secluded DM sector, see, for example, Refs.~\cite{Heikinheimo:2016yds, Tenkanen:2016jic, Arcadi:2019oxh, Lebedev:2021tas}.

The role of inflaton-mediated scatterings in increasing the initial temperature of a chilly relativistic dark sector has been studied in Refs.~\cite{Adshead:2016xxj,Adshead:2019uwj}, where a particularly detailed analysis of the effect of quantum statistics was performed. In this paper, we focus on several other important aspects of this subject, some of which are mentioned in the following:
\begin{enumerate}
\item We consider the complete set of inflaton interactions with the SM gauge, fermion and Higgs fields, by including all the relevant ${\rm SU(3)}_{\rm C}\times {\rm SU(2)}_{\rm L}\times {\rm U(1)}_{\rm Y}$ invariant operators up to dimension $5$. The suppression mass scale for the higher-dimensional operators is chosen in a consistent way, such that the reheat temperature, which is a function of this mass scale, is always lower than this scale. This is necessary for the effective field theory treatment to be valid. 

\item Since we are interested both in the temperature and the number density of the DM particles at later epochs, we solve a coupled system of Boltzmann equations for both these quantities including the full set of relevant inflaton-mediated $2 \rightarrow 2$ collision integrals induced by the above operators (namely, the zeroth and the second moments of the collisional Boltzmann equation for the DM phase-space distribution function). In Refs.~\cite{Adshead:2016xxj,Adshead:2019uwj} it was sufficient for the authors to solve the Boltzmann equation for the relativistic DM energy density, and deduce the temperature evolution from there, since the relic abundance of the DM particles which become non-relativistic at later epochs was not the primary emphasis.

\item We solve for the DM temperature and number density as a function of the DM mass and the DM-inflaton coupling in a very broad parameter space area, in order to determine the values of these parameters that are consistent with the requirements of the total DM abundance, and the constraints from the observations of the CMB anisotropies and the BBN. Thus, we are able to determine the impact of the inflaton-mediated scatterings on the cosmological observables in a detailed way -- which is one of the key new results of our study. We show that the CMB constraints rule out a vast range of the initial DM temperatures for low DM masses, while the BBN bounds exclude a vast range of DM mass values for initial DM temperatures upto an order of magnitude below the SM. Both the CMB and BBN constraints are found to be significantly altered on including the collision effects.

Since the solution of the coupled Boltzmann equations with the multi-dimensional collision integrals over a vast parameter space is already numerically challenging, we did not include the effect of quantum statistics.

\item We point out the possibility that not only can the DM sector be produced with a lower temperature than the SM sector at reheating, it can also be produced with a somewhat higher temperature. The latter scenario can be consistent with the requirement of radiation domination at BBN, since the SM sector can still have a higher energy density owing to its much (two orders of magnitude for a reheat temperature higher than the top quark mass) larger degrees of freedom. This possibility, in fact, leads to a sharp lower bound on the mass of the DM that can saturate the observed DM abundance. For DM masses below this cut-off, the required DM-inflaton coupling is large enough to induce strong back-scatterings, washing out the initial density, which is a novel effect.
\end{enumerate}

With the above goals in mind, we then consider a scenario in which a scalar singlet inflaton field $\phi$ couples to the singlet scalar DM field $\chi$~\footnote{The DM degree of freedom here can be an effective low-energy one, such as a glueball of a hidden non-Abelian gauge group, so that direct couplings to the Higgs field are not induced~\cite{Dolgov:1980uu, Dolgov:2017ujf, Carlson:1992fn, Buen-Abad:2018mas}. If they are elementary fields instead, these renormalizable couplings need to be small enough not to impact the cosmology in our scenario, which also helps in avoiding the direct detection bounds.}, and the SM gauge, Higgs and fermion fields, with the ${\rm SU(3)}_{\rm C}\times {\rm SU(2)}_{\rm L}\times {\rm U(1)}_{\rm Y}$ invariant interaction Lagrangian given by
\begin{align}
\mathcal{L} \supset \mu_\phi \phi H^{\dagger} H + \frac{\lambda_\phi}{2}\phi^2 H^{\dagger} H + \frac{\mu_\chi}{2} \phi \chi^2 + \frac{\lambda}{4} \phi^2 \chi^2+ \frac{1}{\Lambda}\phi\bar{L}He_{R} + \frac{1}{\Lambda}\phi \bar{Q}\tilde{H}u_R +\frac{1}{\Lambda}\phi \bar{Q}Hd_R \nonumber\\ + \frac{1}{\Lambda}(\partial_{\mu}\phi)(g_L\bar{f}_L\gamma^{\mu}f_L+g_R\bar{f}_R\gamma^{\mu}f_R)
+ \frac{1}{\Lambda}\phi B_{\mu\nu}B^{\mu\nu}+\frac{1}{\Lambda}\phi W^{a\mu\nu}W^{a}_{\mu\nu}+\frac{1}{\Lambda}\phi G^{a\mu\nu}G^{a}_{\mu\nu},
\label{eq:Lag}
\end{align}
where,  $\tilde{H}$=$i \sigma_{2}H^{*}$, $H$ being the SM Higgs doublet, $f$ includes all the SM fermions, $L$ and $Q$ are the ${\rm SU(2)}_{\rm L}$ doublet lepton and quark fields, while $e_R, u_R$ and $d_R$ are the ${\rm SU(2)}_{\rm L}$ singlet fields, and $B_{\mu\nu}, W^{a}_{\mu\nu}$ and $G^{a}_{\mu\nu}$ are the field strength tensors for the ${\rm U(1)}_{\rm Y}, {\rm SU(2)}_{\rm L}$ and ${\rm SU(3)}_{\rm C}$ gauge fields, respectively. We have omitted the $CP$-violating interactions for simplicity. Here, the inflaton coupling to the Higgs field is renormalizable with terms of dimension 3 and dimension 4,  whereas the SM gauge invariant couplings to the fermion and gauge bosons are of dimension 5, suppressed by a scale $\Lambda$. Therefore, if the Higgs couplings are appreciable, they will dominate the reheating to the SM sector, as well as the subsequent inflaton-mediated scatterings with the DM particles. On the otherhand, if the Higgs couplings are very small, the dimension-5 terms become relevant. Therefore, in the following, we shall show our results separately for these two scenarios. Although we have considered the reheating of the SM sector through the trilinear scalar interaction in the following, the quartic interaction term $\phi^2 H^\dagger H$ can also be relevant~\cite{Lebedev:2021tas}. 

In what follows, we shall show our primary new results of the impact of inflaton mediated scatterings on the DM temperature and density, and their cosmological implications first in a scenario in which the inflaton dominantly couples to the SM Higgs doublet in Sec.~\ref{sec:sec1}, and then in the complementary scenario in which the inflaton dominantly couples to the SM gauge bosons and fermions through higher dimensional operators in Sec.~\ref{sec:sec2}. 

\section{Scenario-1: Inflaton dominantly couples to the SM Higgs}
\label{sec:sec1}
As mentioned in the Introduction, we first consider the scenario in which the renormalizable inflaton coupling to the SM Higgs doublet is significant, and therefore, we may ignore the dimension 5 couplings to the SM fermion and gauge fields. In this scenario, for perturbative reheating, the relevant decay widths for reheat temperature $T_{\rm R}$ higher than the electroweak phase transition temperature $T_{\rm EW}$, i.e., $T_{\rm R}>T_{\rm EW}$ are given by
\begin{align}
\Gamma_{\phi \rightarrow H^\dagger H} \simeq \frac{\mu^2_\phi}{8\pi m_\phi} \\
\Gamma_{\phi \rightarrow \chi \chi} \simeq \frac{\mu^2_\chi}{32\pi m_\phi} ,
\end{align}
where $m_\phi$ is the inflaton mass. Since the inflaton is considerably heavier than the Higgs and DM particles, we have ignored their masses in the above.
Assuming instantaneous reheating and subsequent thermalization, the initial DM ($T_\chi$) and SM ($T_{\rm SM}$) temperature ratio is given by,
\begin{align}
(T_\chi/T_{SM})_i = g^{1/4}_{*SM}(T_R) \left(\frac{\Gamma_{\phi \rightarrow \chi \chi}}{\Gamma_{\phi \rightarrow H^\dagger H}}\right)^{1/4},
\end{align}
where, $g_{*SM}(T_R)$ is the number of relativistic degrees of freedom in the SM sector at temperature $T_R$. Since we consider only the instantaneous reheating approximation, thermal corrections to the particle masses and the inflaton decay width are not included~\cite{Yokoyama:2005dv}.

The process of internal thermalization for the DM particles produced at reheating is determined by both elastic and inelastic scattering reactions, and following the equilibration of the DM phase-space distribution requires an involved computation~\cite{Ghosh:2022hen,Davidson:2000er}, beyond the scope of the present study. However, we can make an approximate quantitative estimate of the coupling strength necessary for maintaining an internally thermalized DM plasma. 

For very small values of the coupling to the inflaton, the DM sector may not internally thermalize through inflaton-mediated scatterings only. In such cases, as mentioned in the Introduction, additional scalar self-interactions that would expedite the assumed internal thermalization are necessary. The self-interactions for scalar DM arise from renormalizable couplings and therefore may naturally be of appreciable magnitude. 

Let us first consider the role of inflaton-mediated $s-$channel elastic DM self-scatterings in maintaining an internally thermalized DM bath. The $2 \rightarrow 2$ DM self-scattering rate $\Gamma_s$ can be parametrized by
\begin{align}
\Gamma_s &= n_\chi \langle\sigma v\rangle_s = 2 \frac{\rho_\phi}{m_\phi} Br(\phi \rightarrow \chi \chi) \langle\sigma v\rangle_s \simeq  \frac{\rho_\phi}{2 m_\phi} \left(\frac{\mu_\chi}{\mu_\phi}\right)^2 \frac{\alpha^2}{m^2_\phi},
\end{align}
where the $s-$wave thermally averaged self-scattering cross-section has been parametrized as $\expval{\sigma v}_s = \alpha^2/m^2_\phi$, $\alpha$ being the effective self-interaction strength. For the dominant contribution to the $s-$channel inflaton-mediated scatterings, coming from the region around the pole at the centre of mass energy squared $s=m^2_\phi$, we find $\alpha \simeq (\mu_\chi/\mu_\phi)^2$. 
 
The self-scatterings are efficient to maintain an internal thermal equilibrium for $\Gamma_s > \mathcal{H}$, where $\mathcal{H}$ is the Hubble expansion rate, dominated by the inflaton energy density during the reheating phase. The inflaton energy density is approximately determined by the quadratic potential energy in the damped oscillatory phase, $\rho_\phi = \frac{1}{2} m^2_\phi \phi^2$. Hence, the condition for inflaton-mediated self-scatterings being in thermal equilibrium during the inflaton-dominated era is given by,
\begin{align}
\frac{\mu_\chi}{\mu_\phi} \gtrsim 10^{-5} \left(\frac{m_\phi}{10^3 ~\rm GeV}\right)^{1/3}\left(\frac{M_P}{\phi}\right)^{1/6},
\end{align} 
where, $M_P$ is the Planck mass. Now, for high scale inflation with $\phi > M_P$ during inflation, and $\phi \sim M_P$ just after inflation, we can have $\mu_\chi/\mu_\phi \gtrsim 10^{-5}$ for  $m_\phi = 10^3 ~\rm GeV$, as necessary to maintain thermal equilibrium.

For smaller values of $\mu_\chi/\mu_\phi$, scalar DM self-interactions are required for maintaining equilibrium. If we consider $2 \rightarrow 2$ elastic scatterings through an effective $\chi^4$ self-interaction term, they become efficient for 
\begin{align}
\alpha \gtrsim 10^{-15}\left(\frac{\mu_\phi}{\mu_\chi}\right)\left(\frac{m_\phi}{10^3 ~\rm GeV}\right) \left(\frac{M_P}{\phi}\right)^{1/2}.
\end{align} 
With this, even for very small values of  $\mu_\chi/\mu_\phi = 10^{-12}$, one requires a self-interaction strength of $\alpha \gtrsim 10^{-3}$. This implies that the thermalization can be maintained by self-scatterings of DM, with perturbative values of the self-interaction strength. Therefore, we conclude that an internally thermalized DM sector may be realized either through inflaton mediated self-scatterings, or through scalar self-interactions.

As explained earlier, since the inflaton dominantly decays to the SM sector, the reheat temperature can be written as
\begin{align}
T_R \simeq \left(\frac{90}{\pi^2 g_{*SM}}\right)^{1/4} \sqrt{\Gamma_{\phi \rightarrow H^\dagger H} M_P}
\end{align}
It is to be noted that even though the dominant energy density transfer from the inflaton is to the SM sector (namely, a factor of four larger energy transfer for the two trilinear couplings having the same value), there may be situations in which at the time of reheating, the DM temperature is somewhat larger. This is consistent, since the number of relativistic degrees of freedom in the SM sector at these temperatures is around $100$ times larger than the single scalar DM degree of freedom. 

Post-reheating, the DM temperature will evolve due to the inflaton mediated scatterings. It would be convenient to study the evolution of the temperature ratio $\xi = T_\chi/T_{\rm SM}$, which is governed by 
\begin{equation}
\frac{d\xi}{dx}+\frac{\xi}{x}+\frac{\xi}{Y_\chi}\frac{dY_\chi}{dx}=\frac{1}{m_\chi}\expval{\frac{p^4}{3E^3}}+\frac{1}{Y_\chi H s m_\chi}\expval{\frac{p^2}{3E} C[f]},
\label{eq:temp}
\end{equation}
where, $Y_\chi = n_\chi/s$, $H$ is the Hubble rate, $s$ is the entropy density, and the $\expval{...}$ denotes an average over the thermal distribution, see, for example, Ref.~\cite{Ghosh:2022asg} for a derivation. The coupled equation for the scaled DM number density is 
\begin{align}
\frac{d Y_\chi}{dx} =-\frac{s}{H x} \left[\expval{\sigma v}_{\chi \chi \rightarrow H^\dagger H}  (T_\chi) Y_\chi^2 (T_\chi)-\expval{\sigma v}_{\chi \chi \rightarrow H^\dagger H} (T_{SM}) (Y^0_\chi (T_{SM}))^2\right].
\label{eq:yield}
\end{align}
Here, the thermally averaged reaction rate is defined as
\begin{align}
\expval{\sigma v}_{\chi \chi \rightarrow H^\dagger H} (T) = \frac{1}{(n^0_\chi)^2(T)}\int \sigma_{\chi\chi \rightarrow H^\dagger H}~ v ~e^{-\frac{E_1+E_2}{T}}\frac{d^3 p_1}{(2\pi)^3} \frac{d^3 p_2}{(2\pi)^3}.
\label{eq:therm_avg}
\end{align}
For the temperature regime $T>T_{\rm EW}$,  the relevant inflaton-mediated s-channel scattering cross-section is given by the following:
\begin{align}
\sigma_{\chi\chi \rightarrow H^\dagger H} = \frac{1}{8\pi}\frac{\mu^2_\chi\mu^2_\phi}{\sqrt{s(s-4m^2_\chi)}}\frac{1}{(s-m^2_\phi)^2+\Gamma^2_\phi m^2_\phi}.
\end{align}

After electroweak symmetry breaking (EWSB), the $\phi$ and $H$ fields will mix. However, the mixing is very small in the limit $\mu_\phi v << m^2_\phi$, where $v$ is the vacuum expectation value of the Higgs field. We shall be working in this limit in the subsequent analysis. For such small values of the mixing angle with our choice of example parameters, we have checked that the mixing-induced invisible Higgs decay width and spin-independent DM-nucleon scattering rates are below the current experimental bounds. For $T<T_{\rm EW}$, the relevant scatterings would involve a pair of the physical Higgs boson, with the cross-section given by:
\begin{align}
\sigma_{\chi\chi \rightarrow h h} = \frac{1}{32\pi}\frac{\mu^2_\chi\mu^2_\phi}{\sqrt{s(s-4m^2_\chi)}}\frac{\sqrt{1-\frac{4m^2_h}{s}}}{(s-m^2_\phi)^2+\Gamma^2_\phi m^2_\phi}.
\end{align}

For computing the thermally averaged reaction rates, we can simplify the cross-section expressions by using the narrow-width approximation, $\Gamma_\phi << m_\phi$. For example, in one of the representative examples considered, we have chosen the parameter values $m_\phi = 1 ~\rm TeV$, $\mu_\phi = 1 ~\rm GeV$, for which $\Gamma_\phi \sim 4\times 10^{-5}~{\rm  GeV} << m_\phi$. With this approximation, $\sigma_{\chi\chi \rightarrow H^\dagger H}$ in the thermal averaging integral in Eq.~\ref{eq:therm_avg} may be replaced by the following expression
\begin{align}
\sigma_{\chi\chi \rightarrow H^\dagger H} \rightarrow \frac{1}{8\pi}\frac{\mu^2_\chi\mu^2_\phi}{\sqrt{s(s-4m^2_\chi)}}~\frac{\pi}{\Gamma_\phi m_\phi}\delta(s-m^2_\phi).
\end{align}
As we can see from this expression, the maximum contribution to the thermal average reaction rate will come from the region around the $s-$channel pole at $s=m_\phi^2$. On performing the thermal averaging integral, one obtains the well-known form~\cite{Gondolo:1990dk}
\begin{align}
\expval{\sigma v}_{\chi\chi\rightarrow H^\dagger H} (T)= \frac{\mu^2_\chi m_\phi}{32 m^5_\chi} \frac{\pi m_\chi K_1(m_\phi/T)}{T K^2_2(m_\chi/T)}\nonumber\\
\expval{\frac{p^2}{3E}\sigma v}_{\chi\chi\rightarrow H^\dagger H} (T)\simeq \frac{\mu^2_\chi m^2_\phi}{192 m^5_\chi}\frac{\pi m_\chi K_2(m_\phi/T)}{T K^2_2(m_\chi/T)},
\end{align}
where we have dropped the sub-leading terms in the second expression that are numerically found to be small in our case. Considering the appropriate limits for the modified Bessel functions $K_1$ and $K_2$, we obtain the thermal averages for different temperature regimes as follows:
\begin{align}
  \expval{\sigma v}_{\chi\chi\rightarrow H^\dagger H} (T)=
\begin{cases}
    \frac{\mu^2_\chi}{128}\frac{\pi}{T^4},& \text{if } (m_\phi, m_\chi) << T\\
      \frac{\mu^2_\chi}{128}\frac{\pi}{T^4}\sqrt{\frac{\pi m_\phi}{2T}} ~ e^{-m_\phi/T}          &\text{if} ~ m_\phi >> T ,  m_\chi << T
\end{cases}
\end{align}

\begin{align}
  \expval{\frac{p^2}{3E}\sigma v}_{\chi\chi\rightarrow H^\dagger H} (T)=
\begin{cases}
    \frac{\mu^2_\chi}{384}\frac{\pi}{T^3},& \text{if } (m_\phi, m_\chi) << T\\
         \frac{\mu^2_\chi}{768}\frac{\pi}{T^3}\frac{m_\phi}{T}\sqrt{\frac{\pi m_\phi}{2T}} ~ e^{-m_\phi/T}         & \text{if} ~ m_\phi >> T ,  m_\chi << T
\end{cases}
\end{align}
We see from the above expressions that both the thermally averaged collision terms become approximately independent of the inflaton mass  when the bath temperature is larger than the mass. As we shall see subsequently, this effect leads to a substantial impact of the collision processes even for the high-scale inflation scenario, in which both the inflaton mass and the reheat temperature are higher.

\subsection{Numerical Results}
\begin{figure}[htb!]
\centering
\includegraphics[scale=0.55]{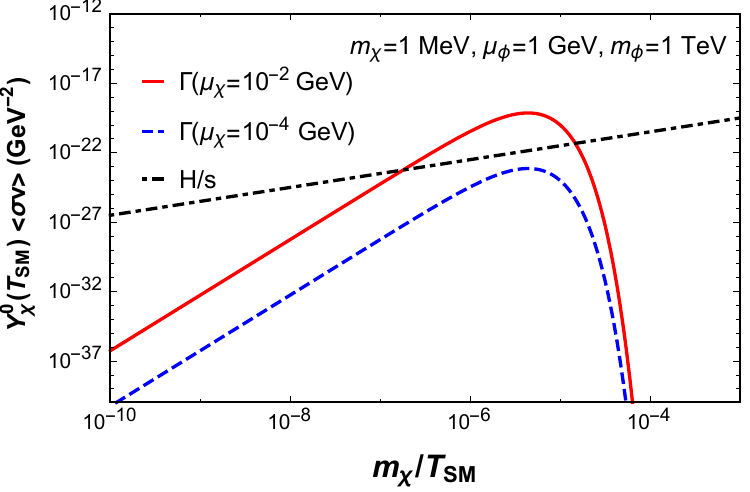}
\includegraphics[scale=0.6]{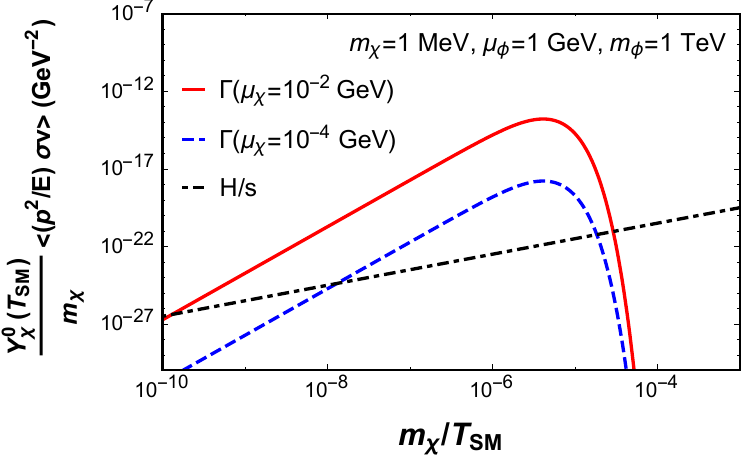}
\caption{\small{\it{Two relevant thermally averaged reaction rates appearing in equations~\ref{eq:temp} and~\ref{eq:yield}, as a function of $m_\chi/T_{\rm SM}$. Also shown is the ratio of the Hubble rate to the entropy density, for comparison.}}}
\label{fig:rate}
\end{figure}

The above features of the thermal averages in the $T_{\rm SM}>m_\phi$ and $T_{\rm SM}<m_\phi$ regions can be seen in Fig.~\ref{fig:rate}, where we have shown the two relevant collision-induced terms appearing in the coupled Boltzmann equations~\ref{eq:temp} and~\ref{eq:yield}, as a function of $m_\chi/T_{\rm SM}$.

\begin{figure}[htb!]
\centering
\includegraphics[scale=0.58]{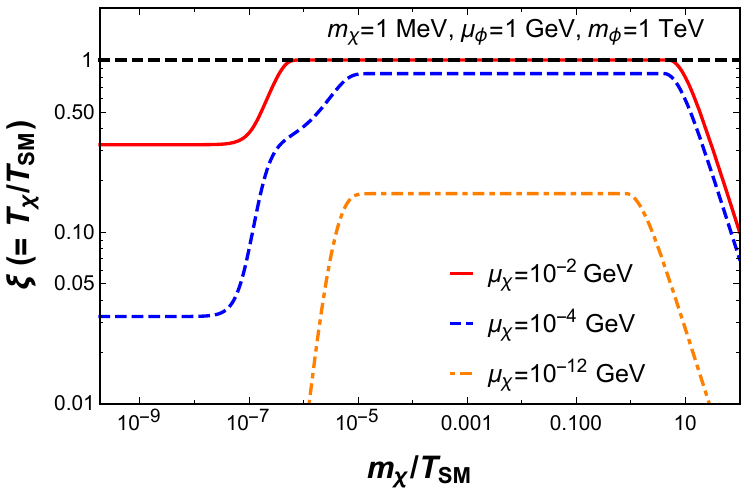}
\includegraphics[scale=0.5]{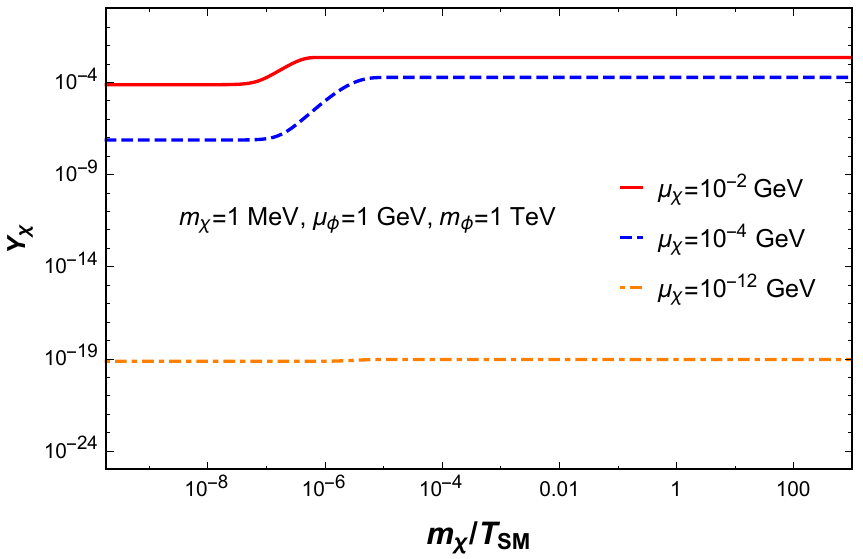}
\caption{\small{\it{Evolution of the DM and SM temperature ratio $\xi = T_\chi/T_{\rm SM}$, as a function of $m_\chi/T_{\rm SM}$ (left), and the corresponding evolution of the DM yield $Y_\chi$ (right). The results are shown for an intermediate reheat temperature of $T_{\rm R} \sim 5 \times 10^6$ GeV and inflaton mass of $m_\phi = 10^3$ GeV.}}}
\label{fig:temp}
\end{figure}

As seen from Eqs.~\ref{eq:temp} and~\ref{eq:yield}, when these rates are comparable to $H/s$, the collision processes can significantly affect the DM temperature evolution. The evolution of the DM and SM temperature ratio is shown in Fig.~\ref{fig:temp} (left column), along with the corresponding DM yields (right column). It is observed that when the collision terms in Eqs.~\ref{eq:temp} and ~\ref{eq:yield} are appreciable, there is a sharp enhancement in $\xi$ by more than an order of magnitude. Even for a coupling $\mu_\chi = 10^{-12}$ GeV, $\xi$ increases significantly,  although for such a small coupling the reaction rates are never above $H/s$. The primary reason for this effect is that for these small values of the DM-inflaton coupling, the initially produced number density of DM at the reheating epoch is small, and the inflaton-mediated DM pair production from SM particle annihilation in later epochs dominates the density. These latter DM particles have an energy comparable to the SM bath temperature, thereby pushing the average DM temperature to higher values. In particular, the temperature ratio $\xi$ changes due to the scattering effects by a factor of $3,26$ and $5\times 10^4$, for $\mu_\chi=10^{-2},10^{-4}$ and $10^{-12}$ GeV, respectively, with the other parameters fixed as $\mu_\phi = 1$ GeV, $m_\chi=1$ MeV and $m_\phi=10^3$ GeV, leading to a $T_R = 5 \times 10^6$ GeV. We also see from this figure that the particle injection through scatterings is effective in the domain $10^{-7}<x<10^{-5}$, with $x=m_\chi/T_{\rm SM}$. Outside this region, $\xi$ remains a constant as long as the DM is relativistic, as expected, and $\xi \propto 1/x$ as the DM becomes non-relativistic.

\begin{figure}[htb!]
\centering
\includegraphics[scale=0.58]{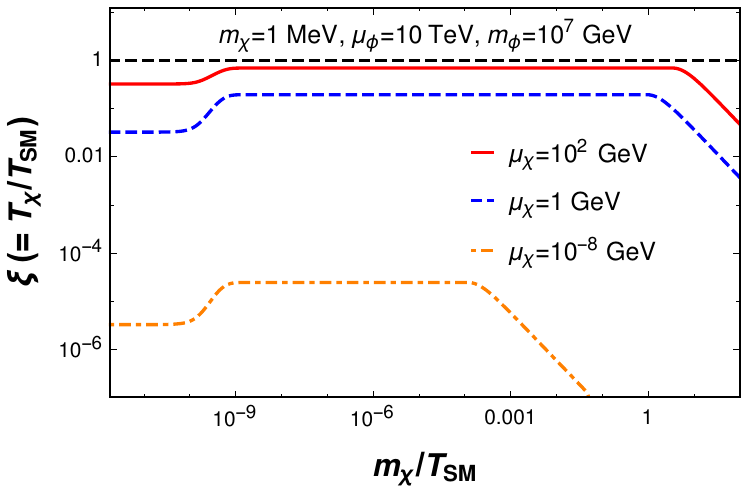}
\includegraphics[scale=0.58]{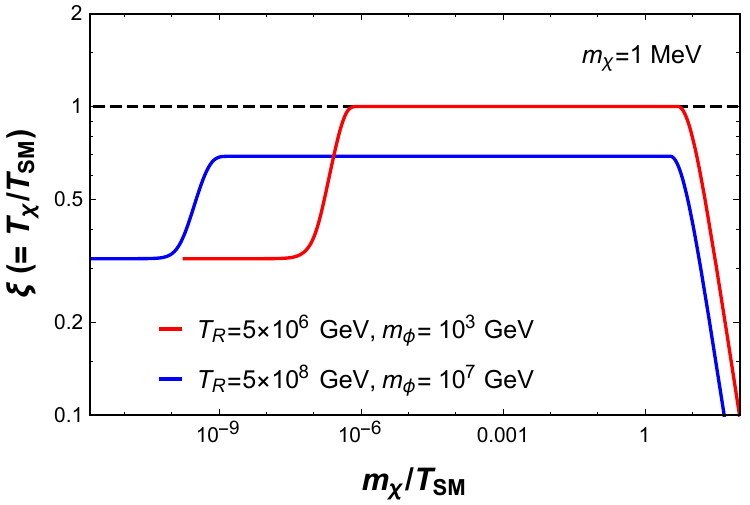}
\caption{\small{\it{Same as Fig.~\ref{fig:temp} (left panel), for a reheat temperature of $T_{\rm R} \sim 5\times 10^8$ GeV and inflaton mass of $m_\phi = 10^7$ GeV (left). Comparison of the evolution of $\xi = T_\chi/T_{\rm SM}$ for two different scales of the inflaton mass and reheat temperature (right), starting from the same initial value $\xi_i$.}}}
\label{fig:temp2}
\end{figure}

The significant modification to the DM temperature is not restricted to the case of an intermediate reheat temperature and inflaton mass, such as $T_{\rm R} \sim 5 \times 10^6$ GeV and $m_\phi = 10^3$ GeV as in Fig.~\ref{fig:temp}. We find that for much larger values of the inflaton mass and the reheat temperature the effects can be important. For example, as seen in Fig.~\ref{fig:temp2} (left panel), for an inflaton mass of around $m_\phi = 10^7$ GeV, considerable changes in the DM temperature can be observed, with $T_{\rm R} \sim 5\times 10^8$ GeV. For comparison, we also show in Fig.~\ref{fig:temp2} (right panel) the evolution of $\xi$ for the two sets of parameters, keeping the ratio of the inflaton couplings to the DM and SM the same, in order to understand the effect of the change in $m_\phi$. This also ensures the same initial value of the temperature ratio $\xi_i$. Thus the effect studied in this paper, which is necessarily present in such reheating scenarios, applies in a broad range of the inflaton mass and reheat temperatures, and hence should be considered in studies of reheat induced DM phase-space properties. The reason behind the impact of the collisions for higher inflaton mass and reheat temperatures, is that, as explained earlier, the thermally averaged collision terms become approximately independent of the inflaton mass  when the bath temperature is larger than the mass. 

\begin{figure}[htb!]
\centering
\includegraphics[scale=0.5]{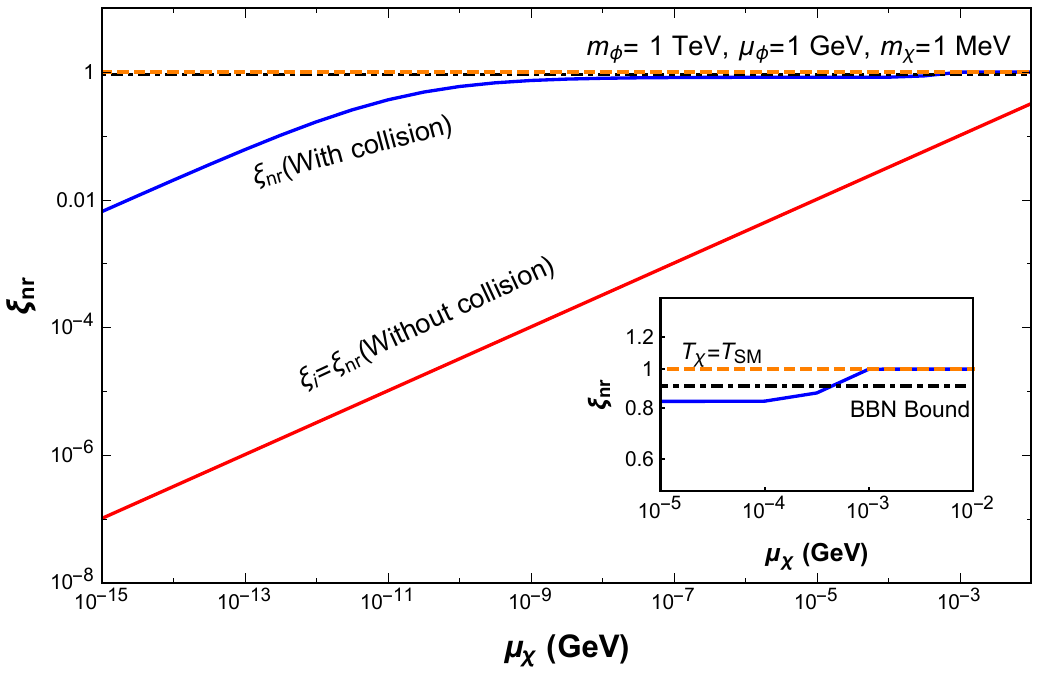}
\caption{\small{\it{Effect of the collisions on the DM temperature in the beginning of its non-relativistic regime, as a function of the DM-inflaton coupling $\mu_\chi$ (left), and a zoomed in version of the figure to illustrate the BBN constraints (inset).}}}
\label{fig:temp_late}
\end{figure}

\begin{figure}[htb!]
\centering
\includegraphics[scale=0.7]{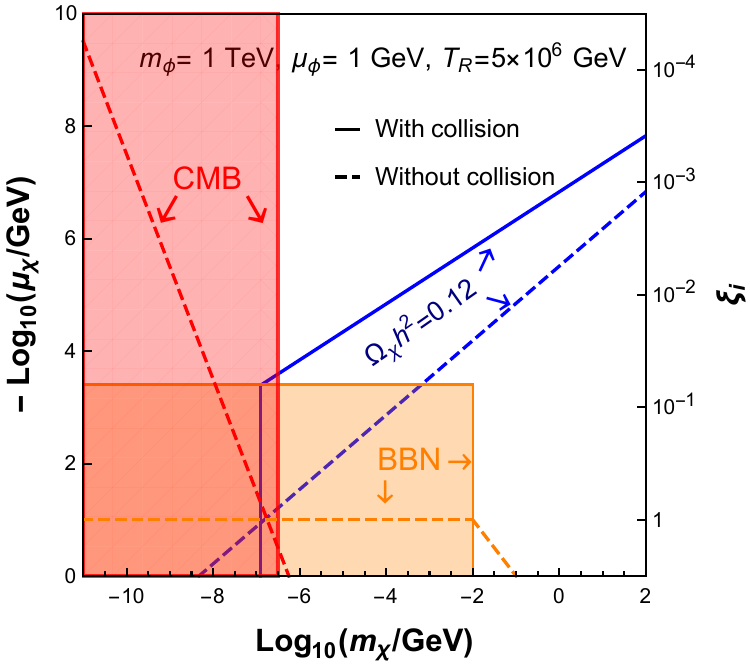}
\caption{\small{\it{Cosmological constraints on the DM mass $m_\chi$ and the DM-inflaton coupling $\mu_\chi$, from considerations of the DM total abundance, the CMB anisotropies and the BBN, both without and with the effect of the collision processes. The initial values of the temperature ratio $\xi_i$ are also shown. Note that the coupling $\mu_\chi$ decreases upward along the $y-$axis in the convention used.}}}
\label{fig:cons}
\end{figure}

The DM temperature at later epochs when it becomes non-relativistic is relevant  from considerations of cosmological observables such as the CMB anisotropies and the matter power spectrum. It may also serve as an important initial condition for number changing processes within the DM sector, if those are present~\cite{Ghosh:2022asg}. Therefore, in Fig.~\ref{fig:temp_late}, we show the effect of the collisions on the DM temperature in the beginning of its non-relativistic regime, as a function of the DM-inflaton coupling $\mu_\chi$. 
In this figure, the non-relativistic value shown $\xi_{\rm nr}$ is defined when $T_\chi = m_\chi/5$~\footnote{The choice of $m_\chi/T_\chi$ at which the non-relativistic regime sets in is somewhat arbitrary, and any order one number gives an estimate. Our choice is motivated from obtaining a smooth analytic approximation to the $\expval{\frac{p^4}{3E^3}}$ thermal average in both the relativistic and non-relativistic regimes.}. As we can see from this figure, for $\mu_\chi > 10^{-3}$ GeV, the effect of collisions forces the DM temperature to be the same as the SM temperature. Therefore for a DM mass of $\mathcal{O} (1 {~\rm MeV})$ the BBN constraints on extra relativistic degrees of freedom should become relevant, as indicated in the zoomed in inset of this figure. The results in Figs.~\ref{fig:rate}-\ref{fig:temp_late} are shown for a specific choice of the DM mass for illustration. The variation of the scattering effects for a wide range of the DM mass, and the corresponding cosmological implications are studied and shown in Fig.~\ref{fig:cons} below for Scenario-1, and in Fig.~\ref{Fig:cons_scen2} for Scenario-2.

Are the values of the DM-inflaton couplings which lead to such a significant effect of the scattering processes on the DM temperature also viable from the point of view of the total DM density? In order to understand this, we show in Fig.~\ref{fig:cons} the contours of $\Omega_\chi h^2 = 0.12$, such that the points on the contours saturate the DM density as determined by the Planck Collaboration~\cite{Planck:2018vyg}, without (blue dashed line) and with (blue solid line) the collision effects. Also shown in the same figure are the constraints in the $\mu_\chi$ and $m_\chi$ plane from the cosmic microwave background anisotropies and the BBN. From the study of CMB anisotropies, one can effectively derive a constraint of $T_\chi(a_{\rm LS})/m_\chi < 10^{-5}$~\cite{Heimersheim:2020aoc}, thus demanding the DM particles to be sufficiently cold in the CMB epoch. Here, $a_{\rm LS}$ is the scale factor at the epoch of last scattering.
The BBN constraint is based on the requirement on the effective number of neutrinos to be $N_\nu = 2.878 \pm 0.278$~\cite{Fields:2019pfx}, thus restricting the contribution of DM particles which are relativistic in the BBN epoch, where we have used the $95\%$ C.L. upper bound on $\Delta N_\nu$ with the SM value being $N_\nu \simeq 3$. 

The inclusion of the collision processes will increase the net abundance of DM through the $s-$channel production from the SM bath, as long as $\xi_i<1$.  Hence, for a given value of the DM mass $m_\chi$, a smaller value of the DM-inflaton coupling $\mu_\chi$ is necessary to saturate the DM abundance, as seen in Fig.~\ref{fig:cons}, by comparing the contours with and without collisions. Note that in Fig.~\ref{fig:cons} the coupling $\mu_\chi$ decreases upward along the $y-$axis in the convention used. The sharp cut-off at around $m_\chi \sim 10^{-7}$ GeV in the relic density contour including collisions is an interesting feature. This indicates that for the choices of $m_\phi=1$ TeV, $\mu_\phi=1$ GeV, which leads to $T_{\rm R}=5 \times 10^6$ GeV, a DM particle lighter than $m_\chi \sim 10^{-7}$ GeV cannot saturate the DM abundance. This feature can be understood as follows. In the relic density contour without collisions we see that for $m_\chi < 10^{-7}$ GeV, one requires a $\mu_\chi \gtrsim 10^{-1}$ GeV to saturate the density, which leads to $\xi_i > 1$. But since both the DM and the Higgs are taken to be effectively massless in this scenario, it is $\xi_i$ which decides the direction of the net energy flow. In particular, for such values of couplings, the DM temperature and yield decrease with scattering, unlike for the $\xi_i < 1$ scenarios seen earlier. 

\begin{figure}[htb!]
\includegraphics[scale=0.58]{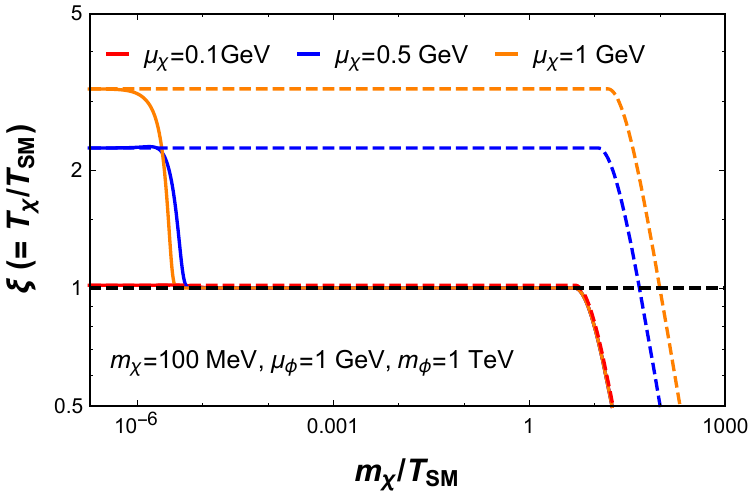}
\includegraphics[scale=0.58]{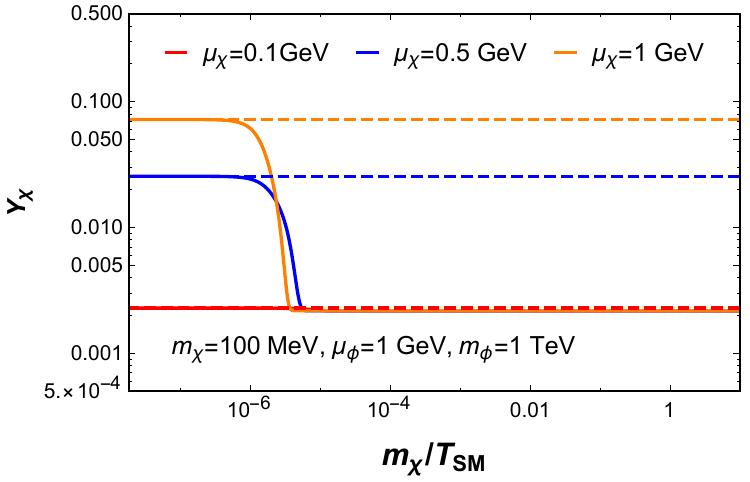}
\caption{\small{\it{The temperature ratio $\xi$ (left) and the DM yield $Y_\chi$ (right) for choices of parameters in which $\xi_i>1$, without (dashed lines) and with (solid lines) the effects of collisions.}}}
\label{fig:large_xi_ini}
\end{figure}

This last feature is further illustrated in Fig.~\ref{fig:large_xi_ini}. In this figure we show the temperature ratio $\xi$ and the DM yield $Y_\chi$ with choices of parameters for which $\xi_i>1$. As explained above, with $\xi_i>1$, the DM temperature decreases with time, and so does the DM yield. In Fig.~\ref{fig:large_xi_ini}, the dashed lines correspond to the scenario in which collision effects are not included, while the solid lines include collision effects. The DM mass chosen here, $m_\chi=100$ MeV would thus have been ruled out by the BBN constraints, as there will be a significant contribution to $\Delta N_\nu$, if there were no collisions, as seen in Fig.~\ref{fig:cons}. However, the effect of collisions decreases the DM temperature for this case, thereby evading the BBN constraints. This explains the difference in the BBN constraints for $m_\chi > 10^{-2}$ GeV and $\xi_i >1$. On the other hand, for $\xi_i<1$, we see the BBN constraints are stronger with collisions, simply because scatterings now increase the DM temperature.

Even if the DM particles are relativistic in the BBN epoch, they are unconstrained by BBN considerations as long as the DM temperature is smaller than the SM one, such that the contribution of DM to the total energy density is insignificant at that epoch. However, although the BBN takes place in a radiation dominated epoch, the CMB decoupling takes place in a matter dominated era. Therefore, if the DM velocity dispersion is large in the CMB era and these DM particles constitute the dominant component of the matter density, there are severe constraints from the observations of the CMB anisotropies. We see this feature in Fig.~\ref{fig:cons}, where for lower masses the CMB observations constrain a very large range of initial $\xi_i$ values. Furthermore, the CMB constraints are significantly affected by the collision processes. For $\xi_i<1$, the inclusion of the collision effects strengthens the CMB bounds, as the DM temperature is pushed to larger values by the scatterings, while for $\xi_i>1$ we see the opposite effect as the DM temperature now goes down due to scatterings. Both these features can be clearly seen by comparing the CMB bounds with (solid line) and without (dashed line) the collision terms in Fig.~\ref{fig:cons}.

\section{Scenario-2: Inflaton dominantly couples to the SM gauge bosons and fermions}
\label{sec:sec2}
We now turn to the scenario in which the inflaton dominantly couples to the SM gauge bosons and fermions, and the reheating to the SM proceeds through these channels. The SM gauge-invariant couplings in this case are of dimension $5$, as in Eq.~\ref{eq:Lag}. Therefore, in order for the effective field theory (EFT) treatment to be valid, we must restrict ourselves to reheat temperatures $T_{\rm R} < \Lambda$, where $\Lambda$ is the scale in which the EFT treatment needs to be replaced with its ultra-violet completion~\footnote{In the scenarios studied in this paper, $T_{\rm R}>m_\phi$, and hence this condition also implies $m_\phi < \Lambda$. Therefore, even though the thermal averages scan a broad range of energy values, their dominant contribution comes from near the $s-$channel pole at $\sqrt{s}=m_\phi < \Lambda$, thereby keeping the effective theory treatment valid.}. This restriction needs to be imposed in a consistent way -- the scale $\Lambda$ also enters the definition of $T_{\rm R}$ through the inflaton decay width. Thus, we solve the parametric equation $T_{\rm R}(\Lambda) < \Lambda$, to determine the consistent set of values for these quantities. For example, with $m_\phi = 10^3$ GeV, this condition requires $\Lambda \gtrsim 10^7$ GeV, and for $m_\phi = 10^7$ GeV, $\Lambda \gtrsim 10^{10}$ GeV. 

The relevant inflaton decay widths to the SM gauge bosons are as follows:
\begin{align}
\Gamma_{\phi\rightarrow i\bar{i}} &= \frac{1}{g_{s}}\frac{1}{2\pi}\frac{m_{\phi}^3}{\Lambda^2} \hspace{2.9cm} [\rm{before\hspace{0.1cm} EWSB}]
\nonumber\\
&= \frac{1}{g_{s}}\frac{1}{\pi}\frac{\sqrt{m_{\phi}^2-4m_i^2}}{\Lambda^2m_{\phi}^2}\Big(
\frac{m_{\phi}^4}{2}-2m_{\phi}^2m_i^2+3m_i^4
\Big)\hspace{0.8cm} [\rm{after\hspace{0.1cm} EWSB}]
\end{align}
where \textit{i} denotes the weak-gauge bosons $W^{\pm},Z^0$, the gluon $g$ and the photon $\gamma$. The symmetry factors are $g_s=2$ for identical particles (\textit{i.e.} for $\gamma,Z^0,g$), and $g_s=1$ for $W^{\pm}$. For the decay width to gluons, there will be an additional colour factor of $8$. 

Before EWSB, the fermionic decay modes of the inflaton are three-body processes, and we have checked that they are numerically subdominant compared to the two-body decay widths to gauge bosons. On the other hand, after EWSB, two-body decay channels to fermion pairs open, with the width given by
\begin{align}
\Gamma_{\phi\rightarrow f\bar{f}} = \frac{1}{16\pi}\frac{\sqrt{m_{\phi}^2-4m_f^2}}{\Lambda^2m_{\phi}^2}\Big(
\frac{v^2}{4}(2m_{\phi}^2-5m_f^2)+8g_{A}^2m_{\phi}^2m_{f}^2\Big)\hspace{0.8cm} [\rm{after\hspace{0.1cm} EWSB}]
\end{align}
where \textit{f} denotes the charged leptons $e,\mu,\tau$, and the up- and down-type quarks $q$, and $g_A=\frac{g_L-g_R}{2}$. We have fixed the coupling factor $g_A=0.5$ throughout the analysis. Neutrinos do not contribute to the above decay width, since in the SM there are no right-handed neutrinos, and the derivative coupling leads to a contribution proportional to $m_f^2$.

The relevant cross-sections for the inflaton mediated $s-$channel scatterings of DM with the SM gauge bosons are given by
\begin{align}
\sigma_{\chi\chi\rightarrow i\bar{i}}&=\frac{1}{g_{s}}\frac{1}{32\pi s}\sqrt{\frac{s}{s-4m_{\chi}^2}}\hspace{0.14cm}\frac{8\mu_{\chi}^2}{\Lambda^2}\frac{s^2}{(s-m_{\phi}^2)^2+\Gamma_{\phi}^2m_{\phi}^2} \hspace{0.9cm} [\rm{before\hspace{0.1cm} EWSB}]\\
&=\frac{1}{g_{s}}\frac{1}{32\pi s}\sqrt{\frac{s-4m_i^2}{s-4m_{\chi}^2}}\hspace{0.14cm}\frac{16\mu_{\chi}^2}{\Lambda^2}\frac{\frac{s^2}{2}-2m_i^2s+3m_i^2}{(s-m_{\phi}^2)^2+\Gamma_{\phi}^2m_{\phi}^2} \hspace{0.9cm} [\rm{after\hspace{0.1cm} EWSB}]
\end{align}
with the symmetry factors and the additional colour factors being the same as for the decay widths above. The corresponding scattering cross-sections to the SM fermions is 
\begin{equation}
\sigma_{\chi\chi\rightarrow f\bar{f}}=\frac{1}{32\pi s}\sqrt{\frac{s-4m_f^2}{s-4m_{\chi}^2}}\hspace{0.14cm}\frac{16\mu_{\chi}^2}{\Lambda^2}\frac{\frac{v^2}{4}(2s-5m_f^2)+8g_{A}^2m_{f}^2s}{(s-m_{\phi}^2)^2+\Gamma_{\phi}^2m_{\phi}^2} \hspace{0.9cm} [\rm{after\hspace{0.1cm} EWSB}].
\end{equation}

\subsection{Numerical Results}
\begin{figure} [htb!]
	\begin{center} 
	    \includegraphics[scale=0.22]{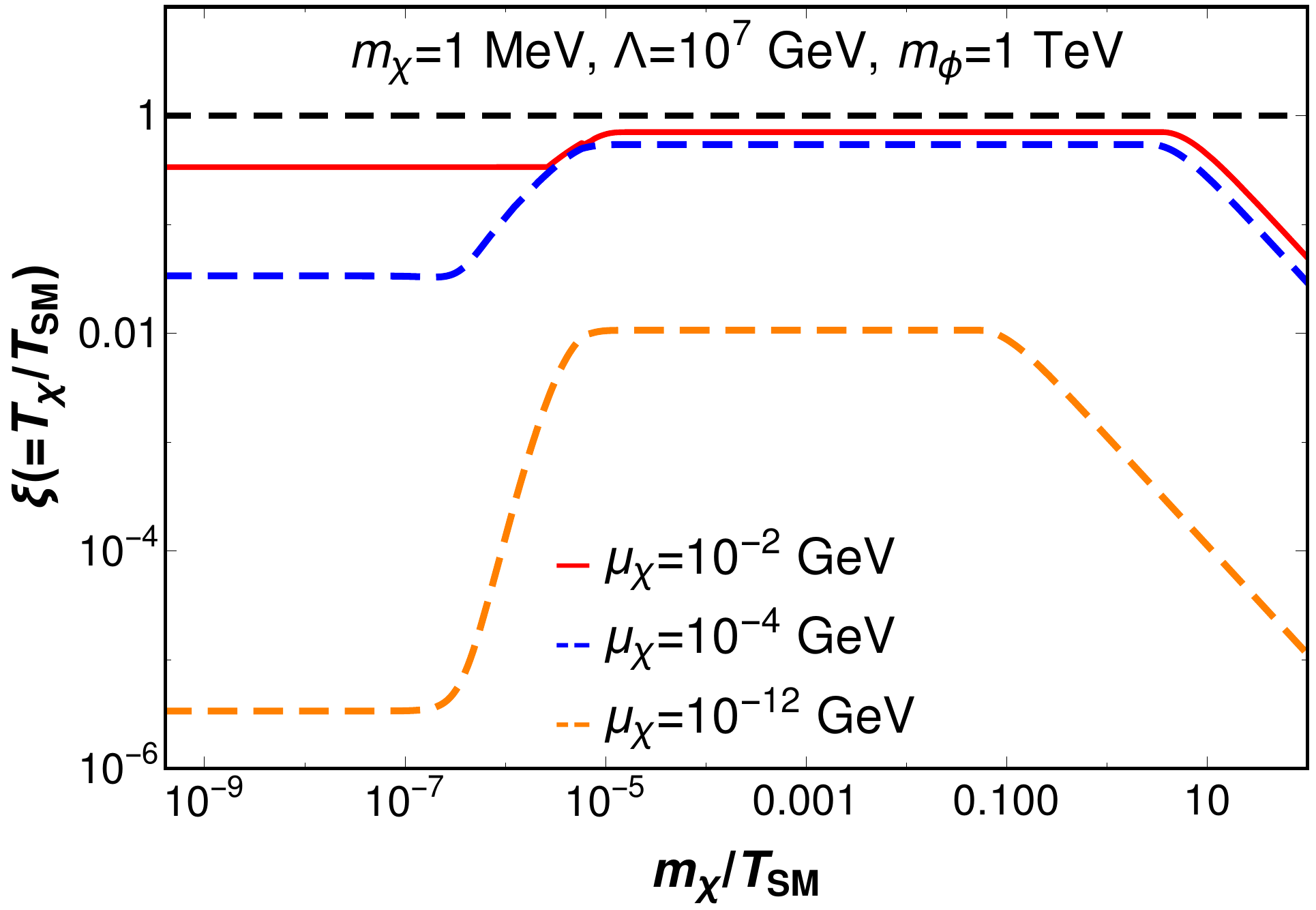}
	    \includegraphics[scale=0.18]{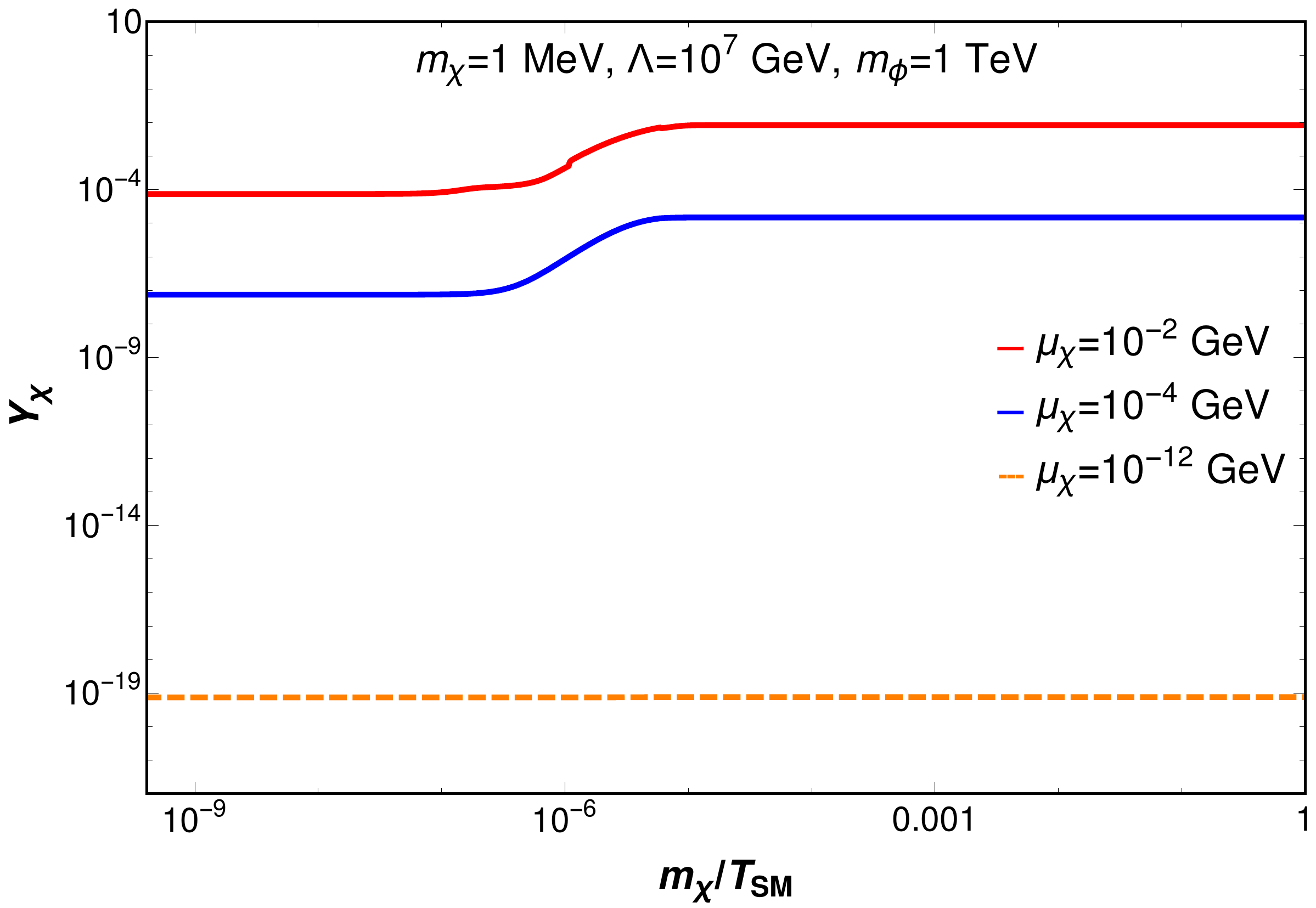}		
	\end{center}
	\caption{\small{\it{Same as Fig.~\ref{fig:temp}, for the scenario in which the inflaton dominantly couples to the SM gauge bosons and fermions. } } }
	\label{Fig:Temperature_and_Abundance}
\end{figure}

Apart from the restriction on the suppression scale $\Lambda$ explained above, and therefore the relatively smaller scattering rates compared to the Higgs scenario, the qualitative features of the results remain the same as in the previous section. In particular, we show in Fig.~\ref{Fig:Temperature_and_Abundance} the evolution of $\xi$ and $Y_\chi$. In this scenario, the temperature ratio $\xi$ changes due to the scattering effects by a factor of $2,15$ and $3\times 10^3$, for $\mu_\chi=10^{-2},10^{-4}$ and $10^{-12}$ GeV, respectively, with the other parameters fixed as $\Lambda = 10^7$ GeV, $m_\chi=1$ MeV and $m_\phi=10^3$ GeV, leading to a $T_R = 2.4 \times 10^6$ GeV. 

\begin{figure} [htb!]
	\begin{center} 
		\includegraphics[scale=0.20]{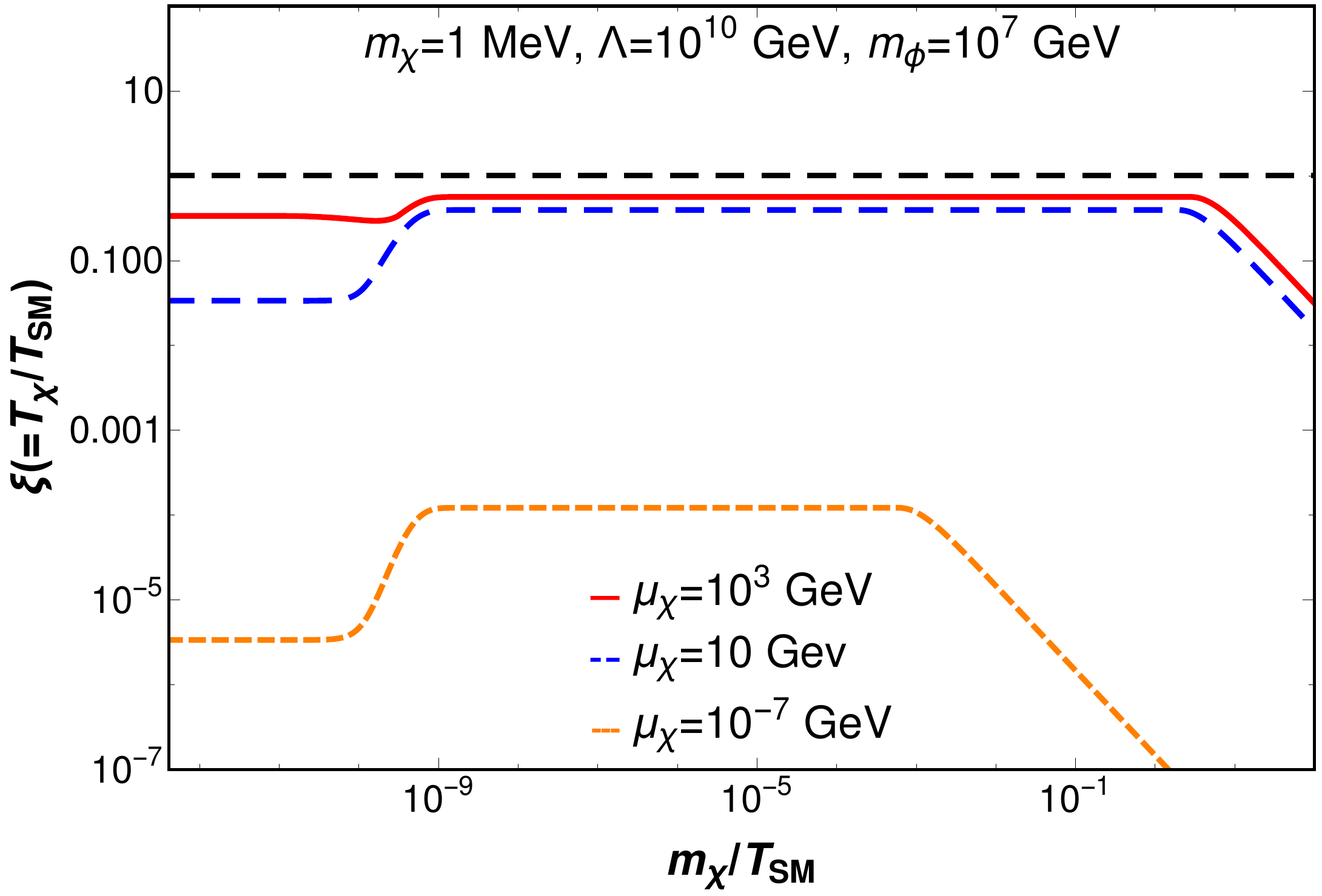}
	    \includegraphics[scale=0.22]{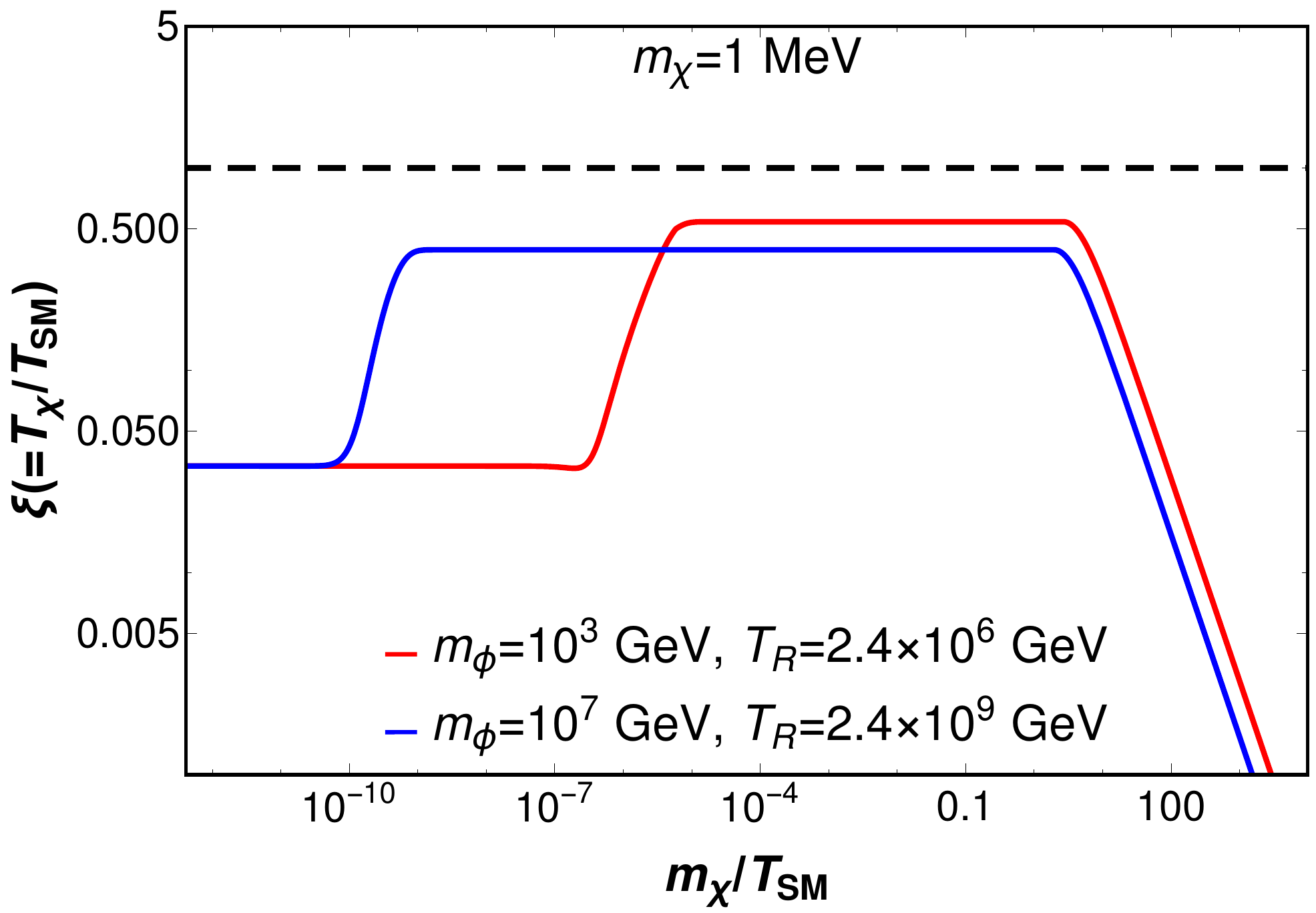}		
	\end{center}
	\caption{\small{\it{Same as Fig.~\ref{fig:temp2}, for the scenario in which the inflaton dominantly couples to the SM gauge bosons and fermions. } } }
	\label{Fig:tem2_scen2}
\end{figure}

The case for a higher reheat temperature and hence a higher suppression scale $\Lambda$ is shown in Fig.~\ref{Fig:tem2_scen2}. We see from this figure that even for a reheat temperature as high as $T_R \sim 2 \times 10^9$ GeV, and an inflaton mass of $m_\phi = 10^7$ GeV, a factor of $10$ change in the temperature ratio $\xi$ is observed. Therefore, this is a significant effect for the high-scale inflation scenario, the reason for which is the same as explained in the previous section. For the right panel in Fig.~\ref{Fig:tem2_scen2}, we have chosen the DM-inflaton coupling such that the initial temperature ratio $\xi_i$ remains the same. For the Higgs coupling scenario, in Fig.~\ref{fig:temp2} (right panel) we see that for $m_\phi=10^3$ GeV, $\xi$ changes approximately by a factor of $3$, while for $m_\phi=10^7$ GeV it changes by a factor of $2$, for the same initial $\xi_i$. Correspondingly, for the gauge and fermion coupling scenario, we see from Fig.~\ref{Fig:tem2_scen2} (right panel) that for $m_\phi=10^3$ GeV, $\xi$ changes approximately by a factor of $16$, while for $m_\phi=10^7$ GeV it changes by a factor of $11$, again for the same initial $\xi_i$. Hence, the relative modification is the same in both cases due to an increase of the inflaton mass. The absolute changes are different, simply because the initial $\xi_i$ values turn out to be smaller in the second scenario due to the higher-dimensional nature of the couplings. 

\begin{figure} [htb!]
	\begin{center} 
		\includegraphics[scale=0.22]{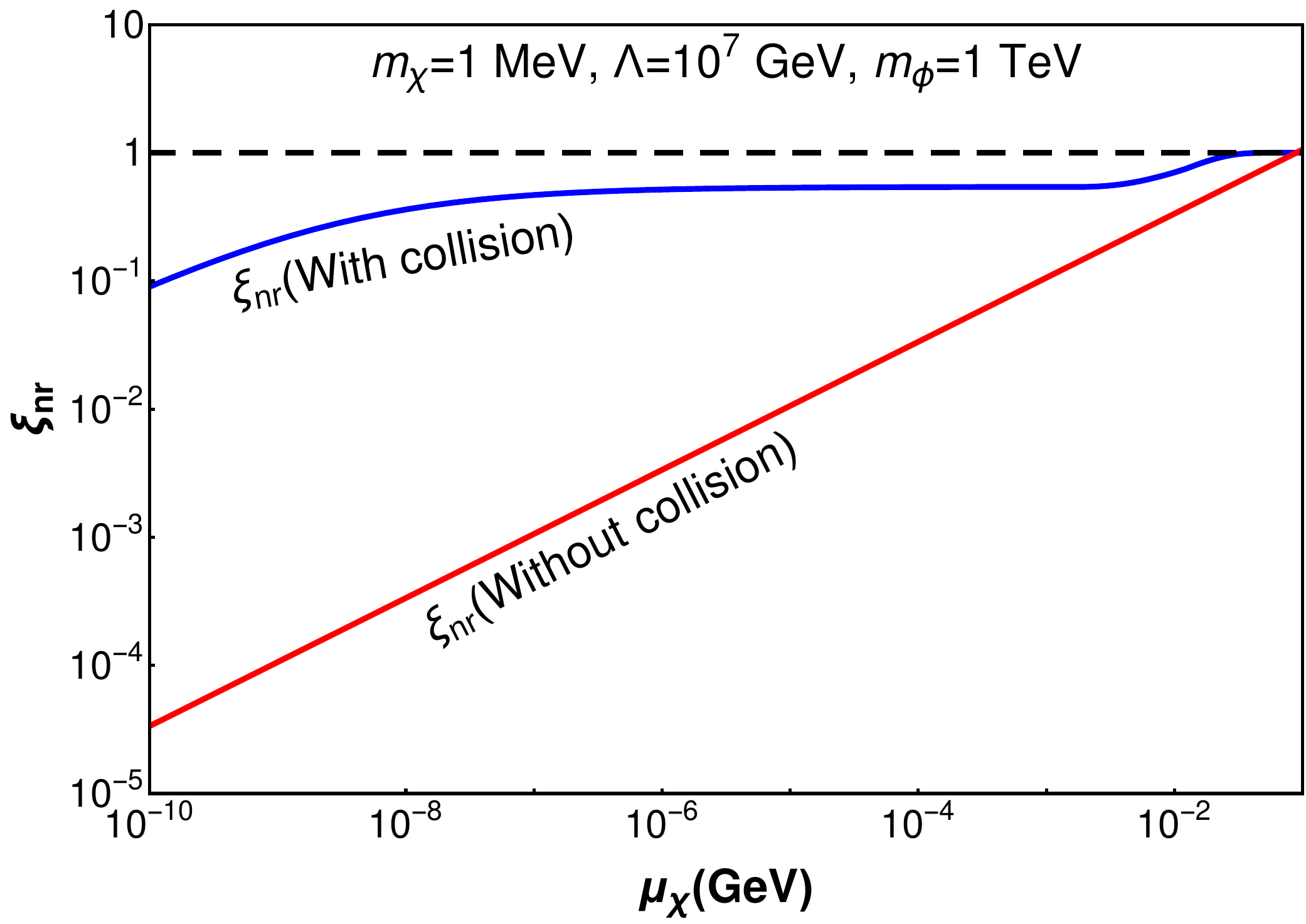}
	\end{center}
	\caption{\small{\it{Same as Fig.~\ref{fig:temp_late} , for the scenario in which the inflaton dominantly couples to the SM gauge bosons and fermions. } } }
	\label{Fig:temp_late_scen2}
\end{figure}

In Fig.~\ref{Fig:temp_late_scen2} we show the effect of the collisions on the DM temperature in the beginning of its non-relativistic regime, as a function of the DM-inflaton coupling $\mu_\chi$, where $\xi_{\rm nr}$ has been defined in the context of Fig.~\ref{fig:temp_late}. Although compared to the Higgs coupling scenario, the effect of collisions is lower with the higher-dimensional couplings, we see from this figure that collisions can modify $\xi_{\rm nr}$ by up to three orders of magnitude for smaller $\mu_\chi$. The difference reduces for larger $\mu_\chi$ values, as the initial DM population produced at reheating dominates.

\begin{figure} [htb!]
	\begin{center} 
	    \includegraphics[scale=0.12]{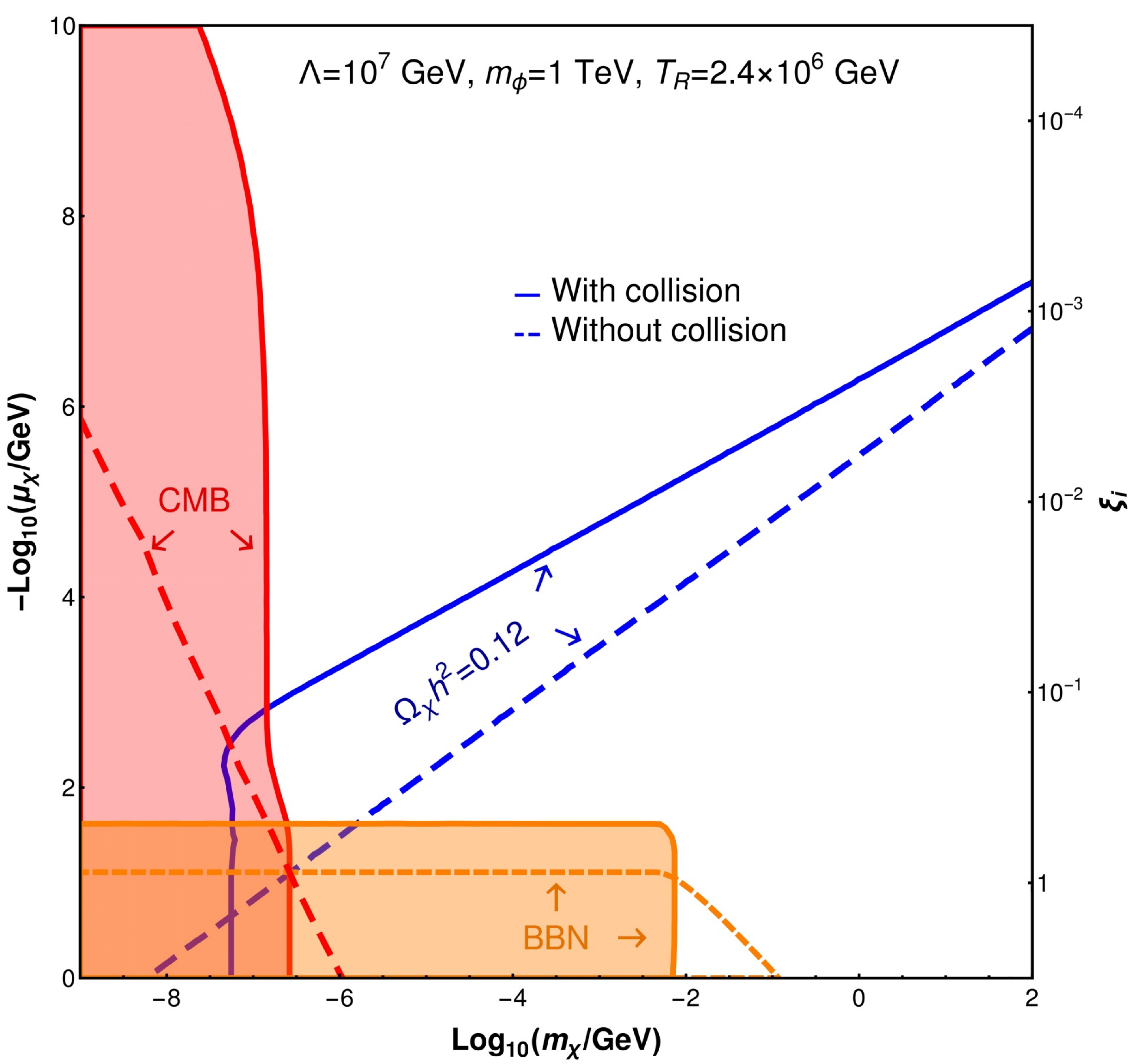}
	\end{center}
	\caption{\small{\it{Same as Fig.~\ref{fig:cons}, for the scenario in which the inflaton dominantly couples to the SM gauge bosons and fermions. } } }
	\label{Fig:cons_scen2}
\end{figure}

Finally, in Fig.~\ref{Fig:cons_scen2} we show the constraints in the $m_\chi$ and $\mu_\chi$ parameter space from the requirements on the total DM abundance, the CMB anisotropies and the BBN, for this scenario. The features observed both without and with collisions are the same as for the inflaton-Higgs coupling scenario, as shown in Fig.~\ref{fig:cons}, and explained in detail in that context. The numerical constraints are somewhat weaker compared to the previous scenario, as the inflaton-SM couplings are suppressed by the higher mass scale. This also leads to the slight weakening of the CMB constraints with collisions for much smaller $\xi_i$ values, unlike for the case in Fig.~\ref{fig:cons}. The sharp lower bound on the DM mass from the requirement of the relic abundance is also seen in this scenario on including the scattering effects.

We thus find that for both the scenarios considered, the inflaton mediated scatterings between the SM and the DM sector can become very significant in determining the DM phase-space distribution and its temperature, leading to important cosmological implications, for a broad range of the inflaton mass and reheat temperature. Therefore, these effects should be included in studies of the DM production and the cosmological constraints on light DM in such reheating scenarios. Our study can be extended to the case in which reheating proceeds through both the perturbative and non-perturbative preheating mechanisms. In particular, thermalization of the DM sector in such scenarios is a non-trivial process, and requires a dedicated separate analysis.

\section*{Acknowledgment}
D.G. thanks Rohan Pramanick for discussions about the numerical computations, and acknowledges the Institute Postdoctoral Fellowship from IISER Kolkata.


\end{document}